\documentclass{ametsoc}

\nolinenumbers
\modulolinenumbers[200]

\journal{jcli}

\bibpunct{(}{)}{;}{a}{}{,}

\title{Indo-Pacific variability on seasonal to multidecadal timescales. \\ Part II: Multiscale atmosphere-ocean linkages}

\authors{Dimitrios Giannakis}

\affiliation{Courant Institute of Mathematical Sciences, New York University, New York, New York, USA}

\extraauthor{Joanna Slawinska\correspondingauthor{Joanna Slawinska, Center for Environmental Prediction, Rutgers, The State University of New Jersey, 14 College Farm Road, New Brunswick, NJ, 08901}} 

\extraaffil{Center for Environmental Prediction, Rutgers, The State University of New Jersey, \\ New Brunswick, New Jersey, USA} 

\email{joanna.slawinska@envsci.rutgers.edu}

\abstract{The coupled atmosphere-ocean variability of the Indo-Pacific on interannual to multidecadal timescales is investigated in a millennial control run of CCSM4 and in observations using a family of modes recovered in Part~I of this work from unprocessed SST data through nonlinear Laplacian spectral analysis (NLSA). It is found that ENSO and combination modes of ENSO with the annual cycle exhibit a seasonally synchronized southward shift of equatorial surface zonal winds and thermocline adjustment consistent with terminating El Ni\~no and La Ni\~na events. The surface wind patterns associated with these modes also generate teleconnections between the Pacific and Indian Oceans, leading to a pattern of SST anomalies characteristic of the Indian Ocean dipole. Fundamental and combination modes representing the tropospheric biennial oscillation (TBO) in CCSM4 are also found to be consistent with mechanisms for seasonally synchronized biennial variability of the Asian-Australian monsoon and Walker circulation. On longer timescales, the leading multidecadal pattern recovered by NLSA from Indo-Pacific SST data in CCSM4, referred to as west Pacific multidecadal mode (WPMM), is found to significantly modulate ENSO and TBO activity with periods of negative SST anomalies in the western tropical Pacific favoring stronger ENSO and TBO variability. Physically, this behavior is attributed to the fact that cold WPMM phases feature anomalous decadal westerlies in the tropical central Pacific and easterlies in the tropical Indian Ocean, as well as an anomalously deep thermocline in the eastern Pacific cold tongue. Moreover, despite the relatively low SST variance explained by this mode, the WPMM is found to correlate significantly with decadal precipitation over Australia in CCSM4.}

\begin{document}

\maketitle

\section{\label{secIntro}Introduction}

The Indo-Pacific is the arena of prominent modes of coupled atmosphere-ocean variability spanning seasonal to multidecadal timescales. Notably, it is the region where the El Ni\~no-Southern Oscillation (ENSO), the dominant mode of interannual climate variability \citep[]{Bjerknes69,Philander90,NeelinEtAl98,WangPicaut04,SarachikCane10,Trenberth13},  takes place. ENSO is well known to exhibit a phase synchronization with the seasonal cycle \citep[][]{RasmussonCarpenter82}, with El Ni\~no and La Ni\~na events typically developing in the boreal summer, peaking during the following winter, and dissipating in the ensuing spring. This phase synchronization has been explained through quadratic nonlinearities in the coupled atmosphere-ocean system producing a cascade of frequencies known as ENSO tones, whose associated temporal and spatial patterns are called ENSO combination modes \citep[][]{SteinEtAl11,SteinEtAl14,McGregorEtAl12,StueckerEtAl13,StueckerEtAl15,RenEtAl16}. In particular, the ENSO combination modes at the sum and difference of the dominant ENSO frequency and the annual cycle frequency project strongly on the surface atmospheric circulation and thermocline depth and are believed to play an important role in restoring El Ni\~no/La Ni\~na conditions to neutral conditions. A prominent feature of ENSO caused by nonlinear dynamics is an asymmetry between El Ni\~no and La Ni\~na phases, which is manifested in positively skewed SST anomaly distributions \citep[][]{BurgersStephenson99,AnJin04,WellerCai13}. 

In addition to ENSO and ENSO combination modes, other modes of interannual Indo-Pacific climate variability which have received significant attention are the Indian Ocean dipole \citep[IOD;][]{SajiEtAl99,WebsterEtAl99} and the tropospheric biennial oscillation  \citep[TBO;][]{Meehl87,Meehl93,Meehl94,Meehl97,LiEtAl02,LiEtAl06}. The IOD is characterized by SST anomalies of opposite sign developing in the eastern and western Indian Ocean, whereas the TBO is generally associated with biennial variability of the Asian-Australian monsoon \citep[][]{Li13}. Similarly to ENSO, the IOD and TBO exhibit a seasonally locked behavior. The existence and significance of linkages between ENSO and these phenomena are areas of active research  \citep[][]{Goswami95,AnnamalaiEtAl04,LiEtAl06,ZhaoNigam15}. 

The Indo-Pacific is also the arena of prominent modes of decadal variability, the most known of which are the Pacific Decadal oscillation \citep[PDO;][]{MantuaEtAl97} and the Interdecadal Pacific Oscillation \citep[IPO;][]{PowerEtAl99}. These two modes are generally thought to be a manifestation of the same phenomenon, and differ mainly in their domain of definition (the PDO is defined over the North Pacific, whereas the IPO is defined over the whole Pacific basin). The physical nature of these modes, and more broadly Pacific decadal variability, has been investigated intensively over the last 20 years, and while a number of hypotheses have emerged, consensus on the validity of these hypotheses is limited. Some studies find that the dominant decadal modes retrieved using classical methods such as empirical orthogonal function (EOF) analysis are merely residuals of ENSO after low-pass filtering; in particular, residuals due to El Ni\~no/La Ni\~na asymmetries resulting from nonlinear dynamics \citep[][]{RodgersEtAl04,Vimont05,SunYu09,WatanabeWittenberg12,SunEtAl14,WittenbergEtAl14}. For instance, \citet{Vimont05} constructs a decadal ENSO-like signal through a linear combination of three interannual modes, with no decadal mode involved, and shows that a low-frequency spatiotemporal residual results from the different spatiotemporal characteristics of these three modes, which can be associated in turn with the ENSO precursor, peak, and ``leftover''. Moreover, \citet{RodgersEtAl04} find that the dominant decadal patterns of SST and thermocline depth, extracted via EOF analysis of low-pass filtered, meridionally averaged, Z20 data over the equatorial Pacific, bear strong similarities to composite patterns representing El Ni\~no/La Ni\~na asymmetries. In light of this finding, they argue that the dominant patterns recovered from decadal low-pass filtered data are a consequence (rather than a cause) of ENSO asymmetries. 

At the same time, other studies associate Pacific decadal variability with  an intrinsic physical signal due to one or more stochastic processes \citep[][]{CaneEtAl95, ThompsonBattisti01, Okumura13}, tropical/ENSO-like dynamics \citep{KnutsonManabe98, YukimotoEtAl00, LiuEtAl02},  or remote forcing from the mid-latitudes \citep{KirtmanSchopf98,BarnettEtAl99,KleemanEtAl99,GuPhilander97,PierceEtAl00,VimontEtAl01,VimontEtAl03,SolomonEtAl03}. Frequently, decadal and multidecadal fluctuations are not associated with a single physical mode, but are considered as multiple phenomena acting simultaneously  \citep{DeserEtAl10, NewmanEtAl16}, or they are viewed as changes in the mean background state \citep[][]{WangRopelewski95}. Such changes may be produced by quasi-decadal variability \citep[][]{HasegawaHanawa06}, global warming \citep[][]{TimmermannEtAl99}, the last glacial maximum \citep[][]{AnEtAl04}, or climate shifts \citep[e.g.,][]{YeHsieh06} that are, most often, attributed to the IPO \citep[][]{SalingerEtAl01,ImadaKimoto09}. Decadal modulations of the tropical biennial oscillation (TBO) have also been studied \citep[e.g.,][]{MeehlArblaster11, MeehlArblaster12}, though less extensively than ENSO. Arguably, progress in elucidating the relation between decadal and interannual variability is hampered by the fact that the extraction of decadal modes through conventional data analysis techniques such EOF analysis often requires ad hoc preprocessing of the data (e.g., low-pass filtering), compounding the physical interpretation of the results.   

Irrespective of its origins and nature, the impacts of the Indo-Pacific multidecadal variability have been documented in numerous studies. \citet{PowerEtAl99} find that ENSO's impact on Australian precipitation depends strongly on the IPO phase, and \citet{SilvaEtAl11} report an IPO modulation of ENSO's impact on South American precipitation. \citet{Dai13} has found a strong correlation of the IPO with anomalously dry and wet decades occurring over the western and central United States in the 20th century. Moreover, multidecadal droughts in the Indian subcontinent and the southwestern United States have also been attributed to the IPO \citep{MeehlHu06}, and the rapid sea level rise/fall observed over the last 17 years over the western/eastern tropical Pacific \citep[called east--west dipole by][]{MeyssignacEtAl12} has been associated with the PDO \citep[e.g.,][]{ZhangChurch12}. \citet{MeehlEtAl13}, \citet{EnglandEtAl11}, and others have associated the decades of global warming hiatus observed during the 20th century with negative phases of the IPO. 

A major challenge in research on decadal variability is the relatively short observational record from which it is difficult to robustly extract modes operating at decadal timescales and to project their activity into the future. \citet{HanEtAl14} examined a longer observational record and showed that the IPO does not explain in itself sea level rise over certain decades. Based on this finding, they hypothesized the existence of another decadal mode. They argued that this mode acts mainly over the tropical Indian Ocean (as opposed to the IPO which is active in the Pacific Ocean), but did not address the question of its origin. \citet{EnglandEtAl14} analyzed the most recent hiatus and associated it with exceptionally strong trade winds (also considered responsible for sea-level rise by \citet{HanEtAl14}) and increased heat uptake by the ocean. Notably, they observed that the magnitude of trade-wind strengthening cannot be explained solely by the IPO, suggesting that other factors, e.g., the internal variability of the Indian Ocean, should play a role. Without further explanation of the origin of trade-wind enhancement, a heat budget analysis of the modeled hiatus was performed based on a simulation with prescribed values of trade winds (as models fail to produce as strong winds as observed; see \citet{LuoEtAl12}). 

Several other studies \citep[][]{KarnauskasEtAl09, SolomonNewman12, SeagerEtAl15} also stress the need to consider modes of Indo-Pacific internal variability other than the IPO to build a more complete explanation of the recent climatic record.  In particular, the second EOF of SST \citep[][]{Timmermann03, YehKirtman04, SunYu09, OgataEtAl13}, the spatial pattern of which consists mainly of an equatorial SST zonal dipole, have been sometimes associated with observed decadal fluctuations. Some authors \citep[e.g.,][]{ChoiEtAl09,ChoiEtAl12,OgataEtAl13} have argued that decadal modulations of ENSO  are driven by this background state. 

In Part~I of this work \citep[][]{SlawinskaGiannakis16}, we constructed a family of spatiotemporal modes of Indo-Pacific SST variability in models and observations using a recently developed eigendecomposition technique called nonlinear Laplacian spectral analysis \citep[NLSA;][]{GiannakisMajda11c,GiannakisMajda12b,GiannakisMajda12a,GiannakisMajda13,GiannakisMajda14,Giannakis15}. A key feature of this method is that it can simultaneously recover modes spanning multiple timescales without pre-filtering the input data. In Part~I, we used NLSA to recover a family of Indo-Pacific SST modes from a 1,300 y integration of the Community Climate System Model version 4 \citep[CCSM4;][]{GentEtAl11} and and a  800 y integration of the Geophysical Fluid Dynamics Laboratory coupled climate model CM3 \citep[][]{GriffiesEtAl11}, as well as from monthly-averaged HadISST data \citep[][]{RaynerEtAl03} for the industrial and satellite eras. This family consists of modes representing (i) the annual cycle and its harmonics, (ii) the fundamental component of ENSO and its associated combination modes with the annual cycle (denoted ENSO-A modes), (iii) the TBO and its associated TBO-A combination modes, (iv) the IPO, and (v) a west Pacific multidecadal mode (WPMM) featuring a prominent cluster of SST anomalies  in the western equatorial Pacific. 

Compared to modes recovered by classical EOF-based approaches, the mode family identified in Part I has minimal risk of mixing intrinsic decadal variability with residuals associated with low-pass filtered interannual modes (since NLSA operates on unprocessed data). Moreover, unlike EOF analysis which mixes distinct ENSO combination frequencies into single modes \citep[][]{McGregorEtAl12,StueckerEtAl13}, the NLSA-based ENSO-A modes resolve individual ENSO frequencies and have the theoretically correct mode multiplicities. In particular, these modes form two oscillatory pairs with frequency peaks at either the sum or the difference between the fundamental ENSO frequency and the annual cycle frequency. In Part I, we saw that the SST anomalies due to the ENSO-A modes are especially prominent in the region west of the Sunda Strait employed in the definition of IOD indices \citep[][]{SajiEtAl99}, indicating that these modes are useful IOD predictors. 

The periodic, ENSO, and ENSO-A modes recovered via NLSA were found to be in good qualitative agreement between the CCSM4, CM3, and HadISST datasets (the latter, for both industrial and satellite eras), and were also verified to be robust against variations of NLSA parameters and spatial domain, as well as measurement noise. The TBO, TBO-A, and WPMM modes (whose corresponding SST anomaly amplitudes are typically an order of magnitude smaller than ENSO) were best recovered in CCSM4, possibly due to the fact that the CCSM4 dataset was the longest studied, although representations of the TBO and modes resembling the WPMM were also found in CM3 and industrial-era HadISST. IPO modes were identified in CCSM4, CM3, and industrial-era HadISST, though in all cases the recovered patterns exhibited some degree of mixing with higher-frequency (e.g., interannual) timescales. The poorer quality of the TBO and decadal modes from CM3 and industrial-era HadISST was at least partly caused by the shorter timespan of these datasets compared to CCSM4, since a similar quality degradation was observed in experiments with subsets of the CCSM4 dataset. We were not able to recover TBO and decadal modes from the satellite-era HadISST dataset.   

In this paper, we employ the mode family from Part~I to study various aspects of coupled atmosphere-ocean variability in the Indo-Pacific on seasonal to multidecadal timescales. First, we examine the phase synchronization of ENSO to the seasonal cycle using spatiotemporal reconstructions of SST, surface winds, and thermocline depth associated with the ENSO and ENSO-A modes. These reconstructions exhibit the southward shift of zonal wind anomalies and recovery of thermocline depth associated with the termination of El Ni\~no and La Ni\~na events in boreal spring \citep[][]{McGregorEtAl12}, and also display traveling disturbances which are consistent with Kelvin and Rossby wave development during ENSO events \citep[][]{StueckerEtAl13}. The reconstructions also exhibit surface winds which are consistent with the Pacific-Indian Ocean SST coupling observed in Part~I. Similarly, we find that precipitation and surface wind reconstructions associated with the TBO and TBO-A modes from CCSM4 are consistent with the biennial variability of the Walker circulation and Asian-Australian monsoon expected from this pattern \citep[][]{MeehlArblaster02}. We also examine El Ni\~no-La Ni\~na asymmetries in NLSA-based representations of ENSO, we find that in both models and observations the corresponding reconstructed SST anomalies exhibit realistic third moments and positive skewness in the equatorial eastern Pacific, so long as a sufficiently complete family of ENSO modes is used in these reconstructions. 

Next, we examine the relationships between the WPMM and the ENSO and TBO modes in CCSM4. We find that the WPMM has sign-dependent modulating relationships with ENSO and the TBO, enhancing the amplitude of the latter during its phase with negative western Pacific SST anomalies. We demonstrate through reconstructions that this cold phase of the WPMM induces anomalous enhanced westerlies in the central Pacific and easterlies in the tropical Indian Ocean, as well as an anomalously deep thermocline depth in the eastern Pacific. Our results are therefore consistent with studies that have shown that such a shallower thermocline profile correlates with strong ENSO activity \citep[][]{KirtmanSchopf98,KleemanEtAl99,FedorovPhilander00,RodgersEtAl04}. We also examine the relationship between the WPMM and decadal precipitation over Australia in CCSM4, and find that the two are correlated at the 0.6 level. In contrast, we do not find such ENSO or precipitation modulation relationships associated with the IPO as extracted from this model via via NLSA.  

The plan of this paper is as follows. In section~\ref{secDataset}, we describe the datasets studied in this work. In section~\ref{secENSOC}, we present spatiotemporal reconstructions of SST, surface winds, and thermocline depth associated with the ENSO and ENSO-A mode families, and discuss aspects of the ENSO lifecycle (including El Ni\~no/La Ni\~na asymmetries) and Pacific-Indian Ocean couplings associated with these modes. We present and discuss analogous spatiotemporal reconstructions for the TBO in section~\ref{secTBO}. In section~\ref{secWPMM}, we present the corresponding reconstructions for the WPMM, and discuss its role in ENSO decadal variability and decadal precipitation over Australia. The paper ends in section~\ref{secSummary} with concluding remarks and a brief outlook. Movies illustrating the dynamical evolution of the NLSA modes are provided as supplementary material.

\section{\label{secDataset}Dataset description}

We study the same Indo-Pacific domain as in Part~I (ocean and atmosphere gridpoints in the longitude-latitude box 28$^\circ$E--70$^\circ$W and 60$^\circ$S--20$^\circ$N), and analyze monthly-averaged data from the same 1300 y control integration of CCSM4 \citep[][]{ccsm4,GentEtAl11,DeserEtAl12b}. In addition to the SST field studied in Part I, we examine surface horizontal and meridional winds, precipitation, and maximum mixed layer depth (XMXL) output  on the respective native atmosphere and ocean grids. In this control integration, both atmosphere and ocean grids have a 1$^\circ$ nominal resolution. We use XMXL as a proxy for thermocline depth.  

We also study monthly-averaged observational and reanalysis data for the satellite era; specifically the period January 1973 to December 2009 studied in Part~I. We use the same monthly-averaged HadISST data \citep[][]{RaynerEtAl03,HadISST13} as in Part~I, and also employ monthly-averaged MERRA \citep[][]{RieneckerEtAl11,merra} and C-GLORS \citep[][]{cglors,StortoEtAl16} reanalysis data for surface winds and the 20$^\circ$C isotherm (Z20), respectively. We use the latter as a proxy for thermocline depth. The HadISST, MERRA, and C-GLORS data are all output on uniform longitude-latitude grids, respectively of $ 1^\circ \times 1^\circ$, $ \frac{1}{2}^\circ \times \frac{2}{3}^\circ $, and $ \frac{1}{4}^\circ \times \frac{1}{4}^\circ $ resolution. For conciseness, we collectively refer to the HadISST, MERRA, and C-GLORS data as observational data. Moreover, when convenient we collectively refer to XMXL and Z20 as thermocline depth data.   
    
Throughout this paper, we work with the NLSA eigenfunctions derived from SST data from CCSM4 and the satellite-era portion of HadISST, which are discussed in sections~4 and~7 of Part~I, respectively. Moreover, all spatiotemporal reconstructions and explained variances are computed as described in section~2c of Part I. Note that we do not use surface wind and thermocline depth data in the computation of the NLSA eigenfunctions. Moreover, unlike Part I, in this paper we do not study observational data for the industrial era, and in what follows all analyses of TBO and decadal modes will be based on CCSM4.

\section{\label{secENSOC}Atmosphere-ocean co-variability of ENSO and ENSO combination modes}

\subsection{Surface wind and thermocline patterns}

\begin{figure}
  \noindent\includegraphics[width=\linewidth]{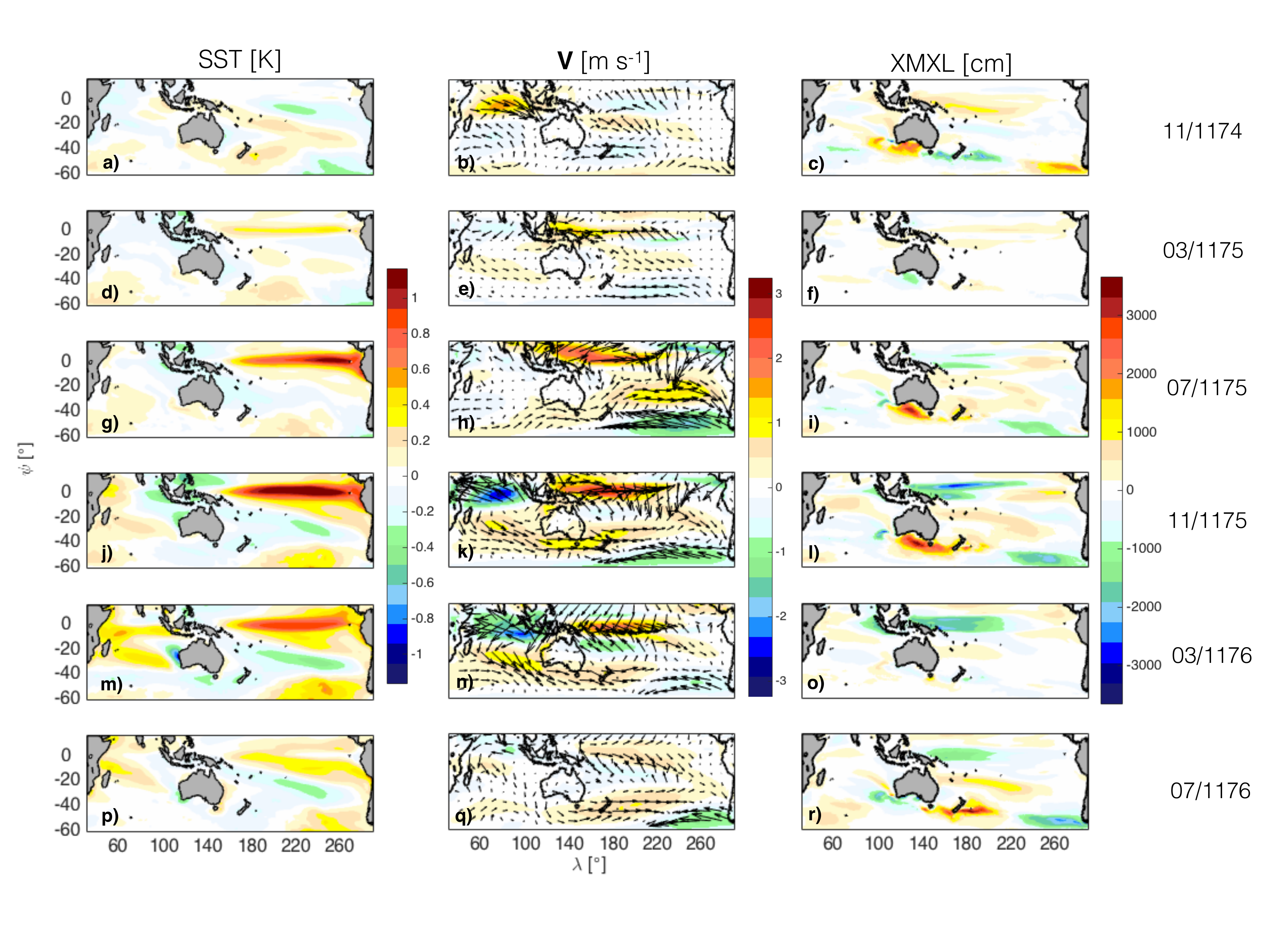}
  \caption{\label{figElNino}Snapshots of reconstructed SST (a, d, g, j, m, p), surface wind (b, e, h, k, n, q), and maximum mixed layer depth (c, f, i, l, o, r) anomalies based on the ENSO and ENSO-A mode families extracted from CCSM4 data via NLSA. The snapshots are taken every four months in the simulation period 11/1174--07/1176, and illustrate the development (a--i), peak (j--l), and dissipation (m--r) of a strong El Ni\~no event taking place in the boreal winter of 1175--1176. Notice the anomalous westerly winds in the western equatorial Pacific (e, h, k) and west-east anomalous thermocline depth gradient (f, i, l) as El Ni\~no conditions develop in March--November 1175, followed by a southward shift of zonal winds (n) and return of climatological thermocline conditions (o, r) as the event decays in February--July 1176. This process is more directly visualized in movie~1 in the supplementary material.}
\end{figure}

\begin{figure}
  \noindent\includegraphics[width=\linewidth]{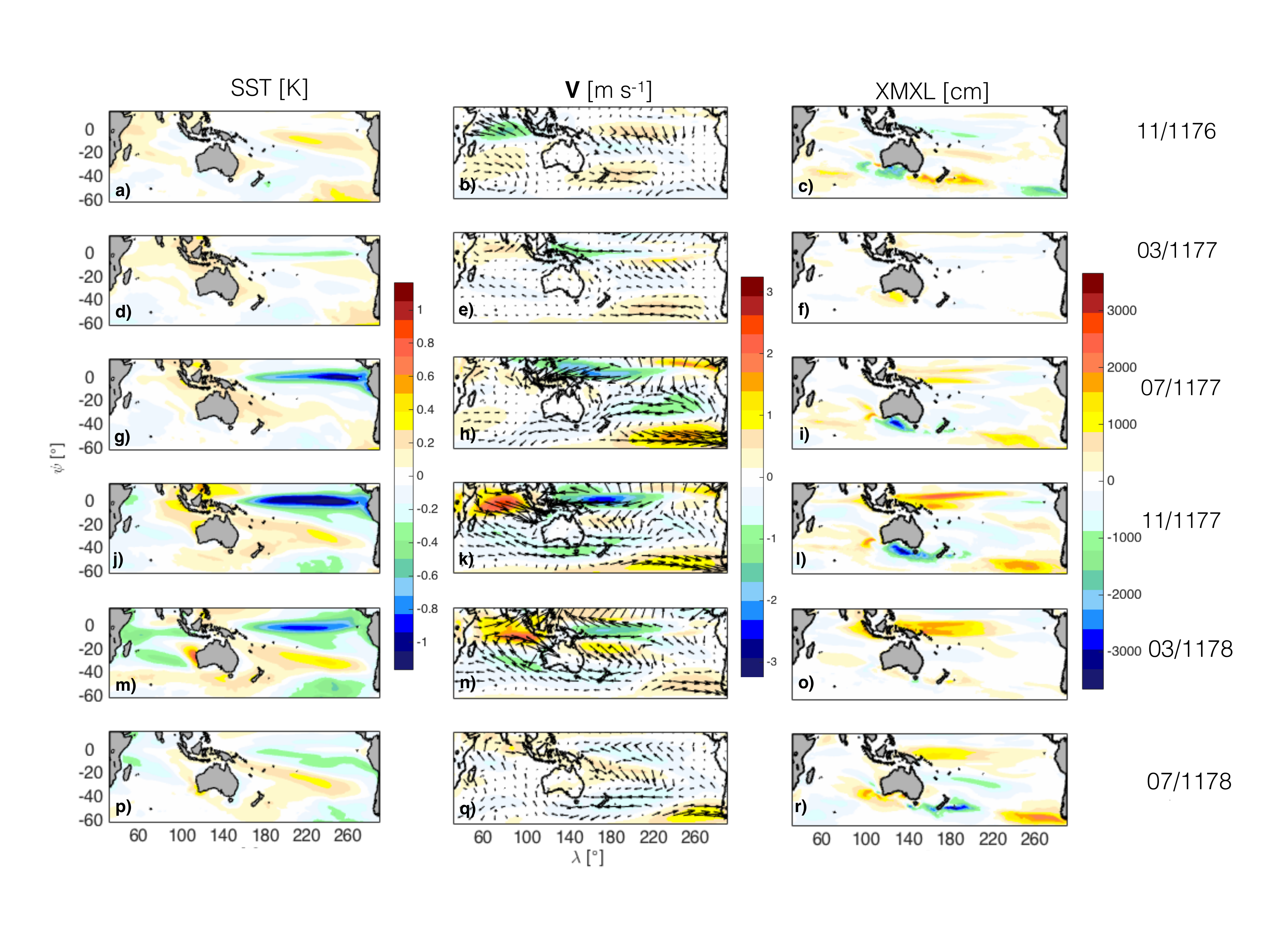}
  \caption{\label{figLaNina}Same as Fig.~\ref{figElNino}, but for the La Ni\~na event of simulation years 1176--1177. (a--c) Neutral conditions in fall of 1176. (d--i) Development of La Ni\~na conditions in spring and summer of 1177, featuring anomalous western Pacific easterlies (h) and anomalous east-west thermocline depth gradient (i). (j--l) Mature La Ni\~na conditions in November 1177. (m--r) La Ni\~na decay in spring and summer of 1178. Notice the southward shift of western Pacific zonal wind anomalies in (k--n).  This process can be more directly visualized in movie~1 in the supplementary material.} 
\end{figure}

\begin{figure}
  \noindent\includegraphics[width=\linewidth]{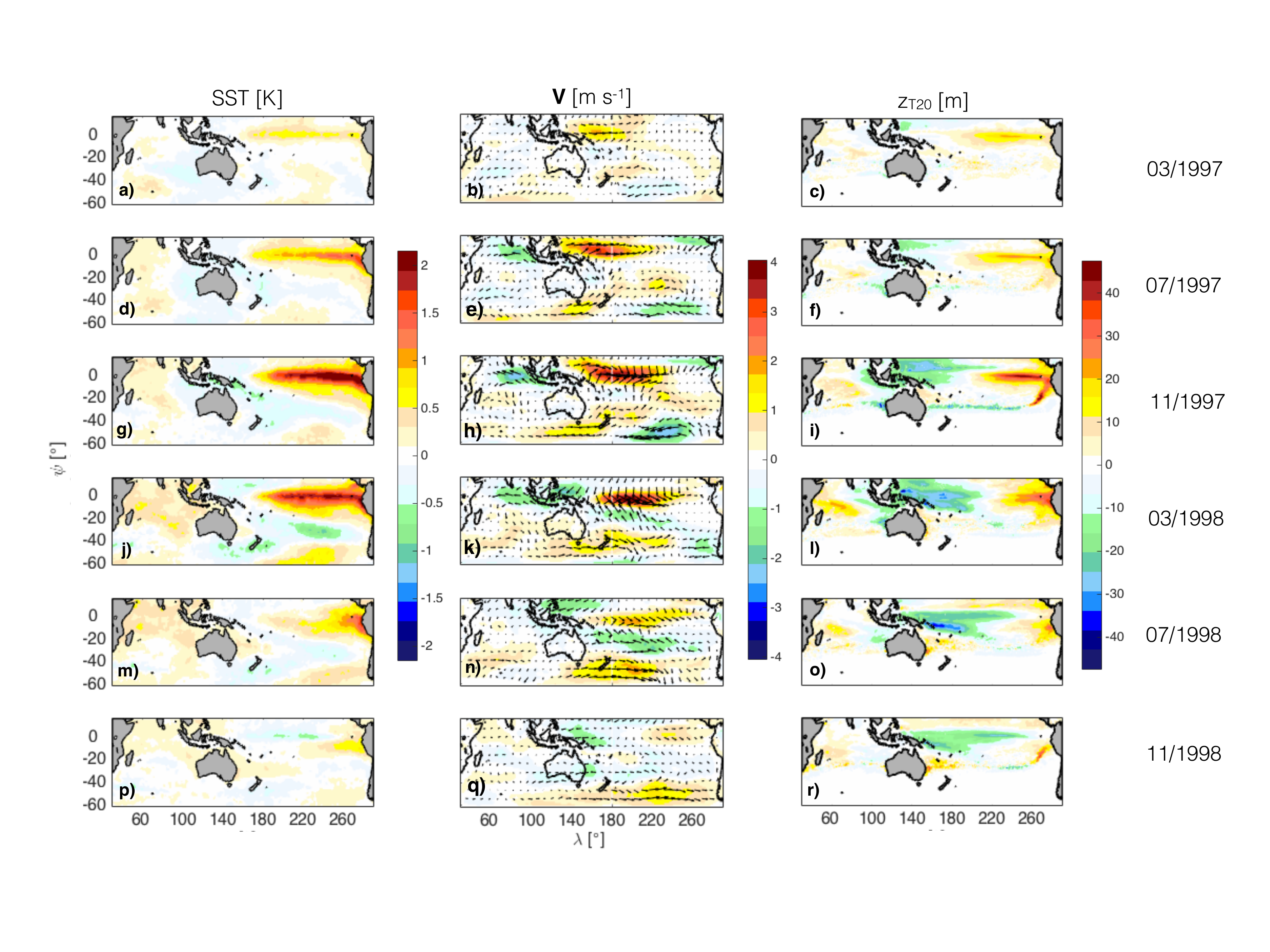}
  \caption{\label{figElNinoH} Same as Fig.~\ref{figElNino}, but for the  El Ni\~no event of 1997--1998 reconstructed using the ENSO and ENSO-A modes recovered from HadISST data via NLSA. (a--f) Buildup of El Ni\~no conditions in boreal spring and summer of 1997. (g--i) Mature El Ni\~no in November 1997. (j--o) Dissipation in boreal spring and summer of 1998. (p--r) Recovery of near-climatological conditions in November 1998. Notice the anomalous easterlies (e, h, k) and east-west thermocline gradient (f, i, l) that develop as El Ni\~no matures. Starting in late fall of 1997 and during the course of El Ni\~no dissipation the anomalous easterlies shift to the south and eventually diminish (h, k, n).  This process can be more directly visualized in movie~2 in the supplementary material.}
\end{figure}

\begin{figure}
\noindent\includegraphics[width=\linewidth]{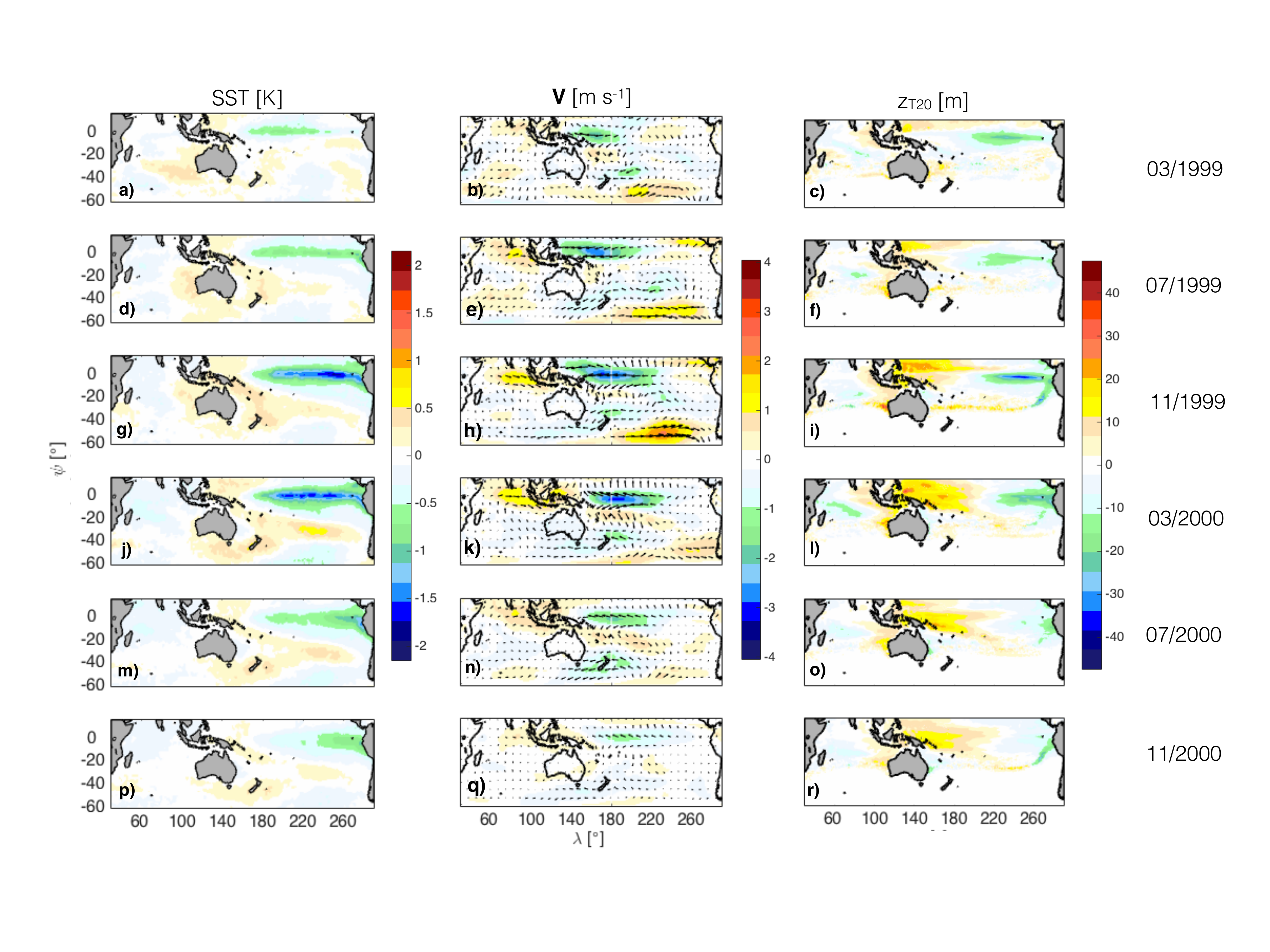}
  \caption{\label{figLaNinaH}Same as Fig.~\ref{figElNinoH}, but for the La Ni\~na event of 1999--2000. (a--f) Buildup of La Ni\~na conditions in boreal spring and summer of 1999. (g--i) Mature La Ni\~na in November 1999. (j--o) Dissipation in boreal spring and summer of 2000. (p--r) Recovery of near-climatological conditions in November 2000.  This process can be more directly visualized in movie~2 in the supplementary material.} 
\end{figure}

It is well known that interannual SST variability in the Pacific Ocean is strongly coupled with surface atmospheric and thermocline variability \citep[][]{AlexanderEtAl02,LengaigneEtAl06,Vecchi06,McGregorEtAl12,StueckerEtAl13,StueckerEtAl15}. In this section, we examine the nature of this coupling in the context of the ENSO and ENSO-A combination modes recovered from the CCSM4 and HadISST data in Part~I. In particular, we examine spatiotemporal reconstructions of SST, surface winds, and thermocline depth associated with this mode family. In what follows, we describe the properties of these reconstructions with reference to specific strong ENSO events in CCSM4 and observations, as we find that individual events are better-suited than composites for describing propagating features from the ENSO and ENSO-A modes evolving on monthly timescales. In the case of CCSM4, we focus on the simulation period 01/1170--12/1180 where ENSO activity is strong (and the WPMM is in its cold phase; see section~\ref{secWPMM}). This period contains a strong El Ni\~no event in the boreal winter of 1175--1176  followed by a strong La Ni\~na event in the boreal winter of 1177--1188. In the case of HadISST, we focus on the period 01/1997--12/2000 containing the strong 1997--1998 El Ni\~no and the subsequent La Ni\~na in 1999--2000. The event from CCSM4 and HadISST are visualized in movies~1 and~2 in the supplementary material, respectively. Representative snapshots from these movies are displayed in Figs.~\ref{figElNino}--\ref{figLaNinaH}. Due to the high temporal coherence of ENSO and ENSO-A modes, the conclusions drawn from these events are representative of other reconstructed ENSO events.

We first consider the reconstructions from CCSM4. As shown in Fig.~\ref{figElNino} and movie~1 in the supplementary material, positive SST anomalies associated with the 1175--1176 El Ni\~no begin building up in the eastern Pacific in FMA 1175. During that period, anomalous westerly winds develop in the western tropical Pacific, and a positive XMXL anomaly travels from the western to eastern equatorial Pacific where it continues to increase in the ensuing months. The eastern Pacific SST anomalies also increase during that time until the event peaks in November-December 1175 reaching $\sim 2 $K SST anomalies in the central-eastern equatorial Pacific. Around the time of the El Ni\~no peak, the anomalous western Pacific westerlies undergo a pronounced southward shift to $\sim 10^\circ$S, which is accompanied by a return of the eastern equatorial Pacific thermocline depth to near-climatological levels by July-August 1176. Meanwhile, the positive SST anomalies in the Eastern Pacific gradually decay until  La Ni\~na conditions begin developing in FMA 1177. The ensuing La Ni\~na episode (see Fig.~\ref{figLaNina}) evolves as a sign-reversed analog of the El Ni\~no event described above.  

More detailed aspects of the thermocline response due to the southward shift of meridional winds is captured in reconstructions in movie~1 in the supplementary material which are based on ENSO-A modes alone. There, it can be seen that as the anomalous westerlies shift to the south of the equator in February-March 1176, a positive XMXL anomaly is generated in the central equatorial Pacific and propagates to the west, reaching the Maritime Continent in August-September 1176 where it contributes to the return of the previously negative XMXL anomaly there to near-climatological levels.  That anomaly is then reflected and begins propagating towards the East, reaching the Eastern Pacific in March--April 1177.  

Overall, the atmosphere-ocean process described above is in good agreement with the predictions of ENSO combination mode theory \cite[][]{SteinEtAl11,SteinEtAl14,McGregorEtAl12,StueckerEtAl13,StueckerEtAl15}. In particular, using intermediate-complexity atmosphere and ocean models, \cite{McGregorEtAl12} propose a mechanism whereby the SST anomalies from a developed ENSO event induce a Gill-type Rossby-Kelvin wave in the lower tropopause which leads to an ageostrophic boundary-layer flow. The latter is stronger in the South Pacific relative to the North Pacific due to weaker climatological wind speeds and associated Ekman pumping in DJF and MAM, giving rise to a seasonally-locked southward shift of meridional winds. Moreover, using a reduced ocean model, they propose that this southward wind shift of meridional winds causes two distinct oceanic responses involving generation of equatorial Kelvin waves or equatorial heat recharge/discharge which are responsible for terminating ENSO events. This seasonally locked southward shift of meridional winds preceding ENSO termination is well captured in Fig.~1 and 2 and movie~1 in the supplementary material. As stated in section~\ref{secIntro} and in Part~I, a representation of the southward zonal wind shift  can also be obtained through the two leading EOFs of atmospheric circulation fields \citep[][]{McGregorEtAl12,StueckerEtAl13}, but this representation mixes the three distinct ENSO and ENSO-A frequencies into a single PC pair. In contrast, the NLSA eigenfunctions recovered from SST isolate the components of ENSO and ENSO-A variability through three distinct pairs of modes evolving at the correct theoretical frequencies. This more detailed and theoretically consistent decomposition should be useful for future studies of ENSO termination.      

Next, we consider the SST, surface wind, and Z20 reconstructions associated with the ENSO and ENSO-A modes extracted from the observational data. As shown in Figs.~\ref{figElNinoH} and~\ref{figLaNinaH} and movie~2 in the supplementary material, SST anomalies preceding the strong El Ni\~no of 1997--1998 begin developing in the equatorial eastern Pacific in DJF 1996--1997 and continue growing until the event's peak in December 1997-January 1998. The buildup of SST anomalies is accompanied by anomalous westerlies in the western Pacific and the establishment of an anomalous shallow (deep) thermocline in the western (eastern) tropical Pacific. The anomalous zonal winds migrate to the south (in a process that begins $\sim 2 $ moths prior to the event peak), reaching a maximum southerly latitude of $ \sim 10^\circ$ S in March 1998, at which point they start to diminish, and the SST and Z20 anomalies in the eastern Pacific decay to climatological levels. ENSO-neutral conditions are established by September 1998, and in December 1997 a La Ni\~na event begins to emerge. The SST anomalies from that event peak in January 2000 and are accompanied by anomalous easterlies in the western Pacific and a negative zonal gradient of Z20 anomalies. The La Ni\~na event rapidly decays in the spring of 2000 following the southward shift and decay of the anomalous easterlies.      

In general, the reconstructions based on CCSM4 and observations are in good agreement over the equatorial Pacific where the majority of ENSO activity takes place---this is to be expected since CCSM4 is known to simulate realistic ENSO variability \citep[][]{DeserEtAl12b}. A Pacific Ocean region where the observational reconstructions differ from those from CCSM4 is the Maritime Continent region north of Sumatra. There, during El Ni\~no events the SST anomalies are positive in the observational data but negative in the CCSM4 data, and an analogous but opposite-sign discrepancy takes place during La Ni\~na events. These discrepancies can also be seen in the composites in Figs.~9--12 in \citet{DeserEtAl12b} which are based on Ni\~no-3.4 indices. 

\subsection{\label{secIOD}Pacific and Indian Ocean couplings}

Besides the tropical Pacific Ocean, the ENSO and ENSO-A modes have strong surface atmospheric and thermocline response in the Indian and South Pacific Oceans. As shown in Fig.~\ref{figElNino} and movie~1 in the supplementary material, coincident with the 1175--1176 El Ni\~no  event is the formation of a prominent anticyclonic surface circulation in the southeast Indian Ocean off the western coast of Australia (see, e.g,. the period December 1175--March 1176). This circulation pattern is consistent with the development of (i) positive SST anomalies in the western Indian Ocean (which are especially pronounced off Africa's east coast) via advection of warm water from the Maritime Continent, and (ii) negative SST anomalies in the eastern Indian Ocean via advection of cold water from the southern Indian Ocean. This dipole SST pattern typically peaks in February, and decays together with the main El Ni\~no SST anomalies during boreal spring. As noted in Part~I, this SST pattern is characteristic of the IOD, and explains $ \sim 35\% $ of the raw nonperiodic variance in the Indian Ocean regions used to define IOD indices \citep[][]{SajiEtAl99}. In the case of La Ni\~na events (Fig.~\ref{figLaNina} and movie~1 in the supplementary material), the southeastern Indian Ocean anticyclone is replaced by a cyclone (centered at the same location), and the positive SST anomalies in the western Indian Ocean are replaced by positive SST anomalies in the eastern Indian Ocean, especially in the region west of the Sunda Strait. 

In the case of observational data, significant Indian-Pacific Ocean couplings also take place during El Ni\~no (Fig.~\ref{figElNinoH}) and La Ni\~na (Fig.~\ref{figLaNinaH}) events (see also movie~2 in the supplementary material), though with some notable differences compared to the CCSM4 data. In particular, unlike the clear dipole SST that develops in the reconstructions from the CCSM4 data during peaking El Ni\~no/La Ni\~na events (see, e.g, December 1175 in movie in the supplementary material), the Indian Ocean SST anomalies are of the same sign in the observational data  (positive and negative during El Ni\~nos and La Ni\~nas, respectively; see, e.g, December 1997 in movie~2 in the supplementary material), though they are generally stronger in the western part of the Indian Ocean (especially east-northeast of Madagascar), giving rise to a zonal SST gradient as in CCSM4. Instead, dipolar SST anomaly patterns with a clear sign reversal over the Indian Ocean can be seen during months preceding El Ni\~no/La Ni\~na peaks; e.g., July 1997 in movie~2 in the supplementary material. 

\subsection{El Ni\~no-La Ni\~na asymmetries}

As discussed in section~5a of Part~I, besides the fundamental ENSO and ENSO-A modes the NLSA spectrum can also contain ``secondary'' ENSO modes with frequencies adjacent to the main ENSO band. These modes form degenerate oscillatory pairs, emerging as the Takens embedding window $ \Delta t$ used in NLSA is increased from interannual to decadal lengths. In particular, the embedding window used to compute modes from the CCSM4 dataset is $ \Delta t = 20 $ y and a number of secondary modes are present in the spectrum; see Fig.~9 in Part~I. This behavior is consistent with NLSA progressively resolving a broadband spectrum (here, due to ENSO) into distinct modes as $ \Delta t $ grows \citep[][]{BerryEtAl13,Giannakis16}. In this section, we examine the role of these secondary ENSO modes in capturing the third-order statistics of ENSO, and in particular the positive SST skewness associated with strong El Ni\~no events \citep[][]{BurgersStephenson99,AnJin04,WellerCai13}. 

\begin{figure}[t]
\noindent\includegraphics[width=\linewidth]{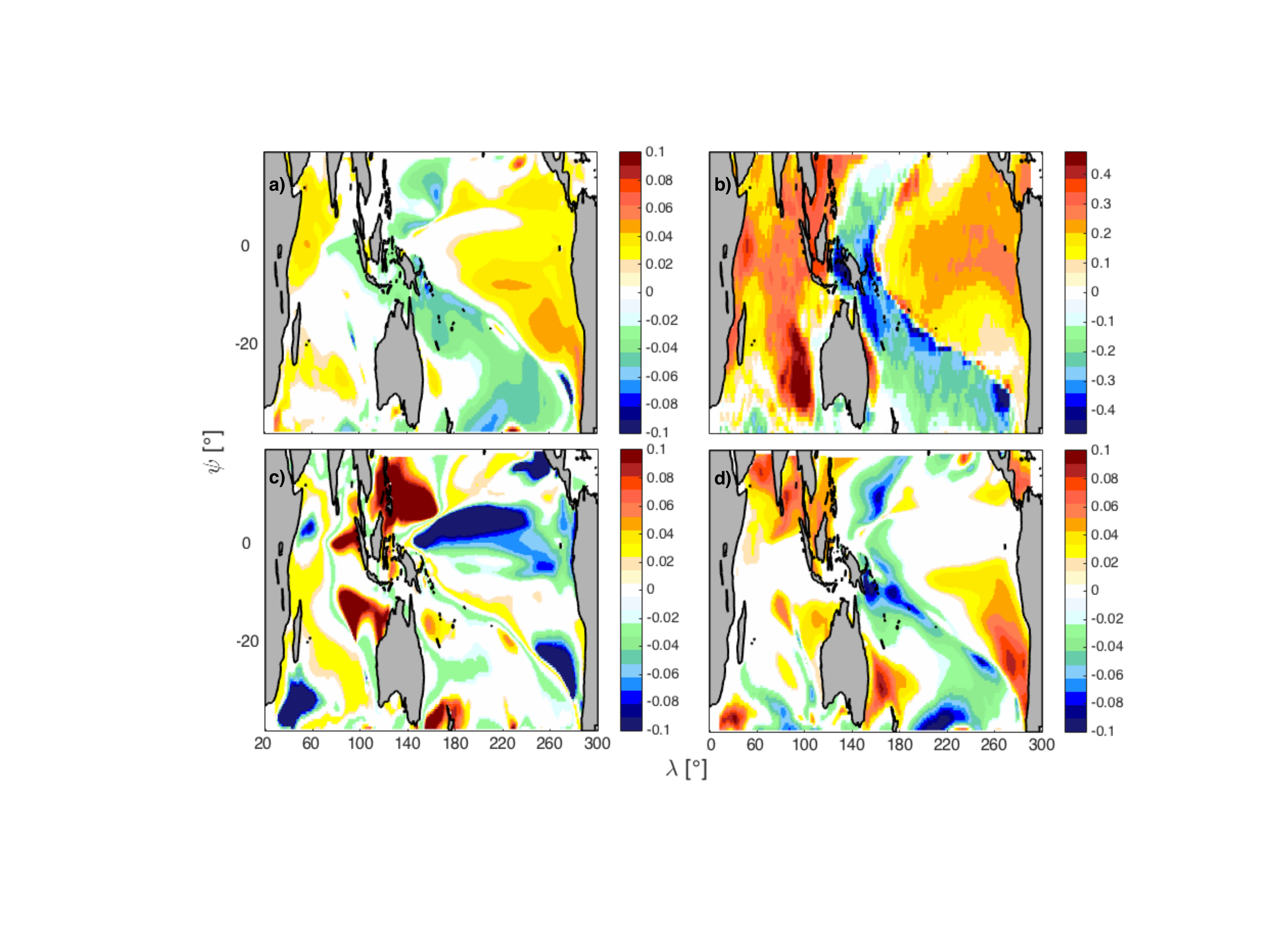}
  \caption{\label{figSkewness}Normalized skewness coefficient maps for reconstructed SST fields from (a, c, d) CCSM4 data using NLSA ENSO modes computed with a 20 y delay-embedding window, (b) industrial-era HadISST data using NLSA ENSO modes computed  with a 4 y delay-embedding window. The modes used in the SST reconstructions are (a) fundamental and secondary ENSO modes and their associated combination modes, (b, c) fundamental ENSO modes and their associated combination modes, (d) secondary ENSO modes and their associated combination modes. The colored gridpoints have statistically significant skewness values based on the test of \citet[][]{White80} with (a, c, d) 15357 and (b) 1665 degrees of freedom corresponding to the number of samples in the NLSA eigenfunctions.}
\end{figure}

Figure~\ref{figSkewness}(a) shows a spatial map of the normalized skewness coefficient $ \gamma_{\text{ENSO},i} $ at each gridpoint $ i $ using the reconstructed SST field obtained from the fundamental ENSO modes and the family of secondary ENSO modes from CCSM4 shown in Fig~9(a) in Part~I, together with their associated combination modes with the annual cycle. The normalized skewness coefficient at gridpoint $i $ is defined as $ \gamma_{\text{ENSO},i} = \overline{ ( y_{ENSO,i} - \bar y_{ENSO,i} )^3 } / \sigma_{ENSO,i}^{3/2}  $, where overbars denote time averages, and $ y_{ENSO,i} $ and $ \sigma_{ENSO,i} $ are reconstructed SST fields and their standard deviations for the selected mode family (see section~2c in Part~I). In Fig.~\ref{figSkewness}(a), skewness is positive over the eastern Pacific cold tongue region associated where the majority of ENSO activity occurs, and takes negative values in the western Pacific. Moreover, the Indian Ocean features a dipolar skewness pattern with predominantly positive (negative) values in the western (eastern) parts of the basin. These features are  consistent with skewness patterns constructed from detrended observational SSTs \citep[see, e.g., Fig.~6(c) in][]{WellerCai13} (although the skewness values in Fig.~\ref{figSkewness}(a) are about a factor of 10 smaller than the corresponding values from raw SSTs.) Using more secondary ENSO modes, or a shorter embedding window as illustrated in Fig.\ref{figSkewness}(b), leads to higher skewness values from those shown in Fig.~\ref{figSkewness}(a).   

Next, to assess the contributions of the individual modes in the ENSO family used to build the map in Fig~\ref{figSkewness}(a), we examine skewness maps based on either the fundamental ENSO and ENSO-A modes (Fig.~\ref{figSkewness}(c)), or the secondary ENSO and ENSO-A modes (Fig.~\ref{figSkewness}(d)) alone. Interestingly, the skewness maps based on the fundamental modes take negative values in the eastern Pacific cold tongue, and those based on the secondary modes are positive but much weaker there. These facts indicate that the positive skewness from the full ENSO family is an outcome of product terms between the fundamental and secondary ENSO modes. Assuming that the secondary ENSO modes are part of a ENSO frequency cascade \citep[][]{StueckerEtAl13} resolved by NLSA (as conjectured in section~5a of Part~I), these results suggest that at least in CCSM4, the positive eastern Pacific SST skewness due to ENSO may be the outcome of an interaction between the fundamental and secondary frequencies in that cascade.        

In the case of HadISST data, the modes were computed using a 4-year embedding window (see section~7 in Part~I), and as a result a single set of ENSO and ENSO-A modes represents the ENSO variability that would be split into fundamental and secondary modes at longer embedding windows. As shown in Fig.~\ref{figSkewness}(e), the normalized skewness map computed from the ENSO and ENSO-A family also recovers the basic features seen in skewness maps from raw SST data, and moreover exhibits generally higher values than the model data.

\section{\label{secTBO}Atmosphere-ocean co-variability of the TBO and TBO combination modes in CCSM4}

Next, we study SST, surface wind, and precipitation patterns associated with the TBO and TBO-A modes recovered from CCSM4. As in section~\ref{secENSOC}, we discuss the evolution of this patterns with reference to a specific TBO event occurring in the simulation period 12/1169--12/1171. Snapshots from this period are displayed in Fig.~\ref{figTBO}, and movie~3 in the supplementary material shows the full dynamical evolution of the SST, surface wind, and precipitation fields associated with the TBO and TBO-A modes. 

As shown in Fig.~\ref{figTBO}(a--c), the boreal winter of simulation year 1169 is an anomalously weak Australian monsoon year featuring horizontal surface divergence (subsidence) and negative SST anomalies over the western Pacific, and negative precipitation anomalies stretching south from the Maritime Continent to northern Australia. This warm pool divergence drives anomalous easterlies over the eastern and central Indian Ocean and anomalous westerlies over the central tropical Pacific. The anomalous easterlies correspond to a weakened western Walker Cell, enabling the development of positive SST anomalies in the western Indian Ocean and intensification of deep convection (as evidenced by positive precipitation anomalies in east Africa). \citet[][]{MeehlArblaster02} associate this pattern with a Rossby wave response over Asia and positive temperature anomalies over the Indian subcontinent, preconditioning for a strong Indian monsoon in the upcoming boreal summer. Indeed, in JJAS 1170 an anomalously strong Indian monsoon takes place (see positive precipitation anomalies over the Indian subcontinent and Southeast Asia in Fig.~\ref{figTBO}(f)). During that time, the western Walker cell is anomalously strong (Fig.~\ref{figTBO}(e)), and anomalous precipitation occurs over the tropical eastern Indian Ocean and western Pacific driven by positive SST anomalies there (Fig.~\ref{figTBO}(d)). \citet[][]{MeehlArblaster02} attribute the development of these anomalies to thermocline adjustment resulting from eastward-propagating Kelvin waves generated by the anomalous circulation in the preceding winter and spring months. 

The positive SST anomalies in the waters off Australia and the Maritime Continent persist over the ensuing boreal fall and winter (Fig.~\ref{figTBO}(g)). As a result,  as precipitation gradually shifts southeastward from India and Southeast Asia in response to the annual cycle of insolation, it becomes anomalously strong. This results in an anomalously strong Australian monsoon in the boreal winter of 1170--1171 (Fig.~\ref{figTBO}(i)),  completing the first half of the TBO cycle. In 1171--1172, the mechanism described above is effectively sign-reversed, and an anomalously week Indian Monsoon takes place (Fig.~\ref{figTBO}(j--l), followed by a weak Australian monsoon in the boreal winter of 1172--1173 (not shown), completing the TBO cycle.   

In summary, we find that in CCSM4 the spatiotemporal SST, precipitation, and surface wind patterns represented by the TBO and TBO-A modes are broadly consistent with the mechanism  for biennial monsoon variability proposed by \citet[][]{MeehlArblaster02}. It is important to note that as with ENSO, both the fundamental and combination (TBO-A) modes are essential in representing this seasonally-locked process, but due to the weak ($\sim 0.1$ K) SST anomalies involved, these modes are challenging to capture from unprocessed observational data. In Part~I we saw that it was possible to recover via NLSA modes resembling the TBO (albeit with mixing between the fundamental and combination frequencies) from HadISST data spanning the industrial era, but not the satellite era for which reliable reanalysis data are available. Nevertheless, longer historical datasets are available for certain variables relevant to the TBO such as Indian rainfall \citep[][]{RajeevanEtAl06}, and it should be an interesting topic for future work to study such datasets using the modes recovered here from industrial-era HadISST in studies of the TBO.       

\section{\label{secWPMM}Decadal-interannual interactions and climatic impacts in CCSM4}

\subsection{Amplitude modulation of ENSO and the TBO}

In this section, we examine the modulating relationships between the WPMM and other interannual and decadal modes in the NLSA spectrum. Figure~\ref{figHilbert} displays the temporal pattern corresponding to the WPMM, together with the amplitude envelopes of the fundamental ENSO modes,  the TBO modes, and the IPO mode recovered from the CCSM4 data (see section~4 in Part~I). There, one can clearly notice that the envelopes of the fundamental ENSO modes and the biennial modes correlate negatively and significantly with the temporal pattern of the WPMM. In particular, the correlation coefficients equal $-0.63$ and $-0.52$, respectively for ENSO and the TBO. Moreover, it is evident that the nature of these modulating relationships is sign-dependent, in the sense that strengthening of the interannual modes occurs predominantly during negative phases of WPMM, i.e., during negative SST anomalies in the western tropical Pacific. On the other hand, the correlations between the amplitudes of the IPO-like mode and the WPMM are significantly weaker (with a correlation coefficient of $-0.11$; see Fig.~\ref{figHilbert}(c)). Similarly weak correlations are also observed between the IPO-like modes and the amplitudes of the dominant ENSO and biennial modes. 

We note that in the case of the modes recovered from CM3 (see section~6 in Part~I), we also observe amplitude modulation relationships  (not shown here) between the WPMM and the interannual modes, but these relationships are somewhat different than in CCSM4. In particular, we find that in CM3 the negative phase of the multidecadal fool correlates with increased ENSO and TBO amplitude, but during the positive WPMM phase there is no significant ENSO and TBO amplitude suppression as observed in CCSM4. As a result, the corresponding time-averaged correlation coefficients are about a factor of two smaller than the CCSM4 results in Fig.~\ref{figHilbert}. Another contributing factor in this decrease of correlation is that the  temporal patterns of the WPMM and TBO in CM3 are noisier than  in CCSM4.   

To physically account for the amplitude modulation role of the WPMM, we examine its associated reconstructed SST, surface wind, XMXL patterns, and precipitation patterns. As with ENSO and the TBO, we discuss these patterns with reference to a specific event, namely a cold WPMM event taking place during simulation years 1165--1185 (note that this interval includes the intervals employed in sections~\ref{secENSOC} and~\ref{secTBO} in our discussion of ENSO and TBO, respectively). The 1165--1185 cold WPMM event is illustrated through composites in Fig.~\ref{figWPMM}, and can be directly visualized in movie~3 in the online supplementary material.  There, it is evident that during the WPMM's cold phase, the decadal cluster of negative SST anomalies in the West Pacific is collocated with surface wind divergence, leading to anomalous westerlies in the central and eastern Pacific and anomalous easterlies in the western Pacific region north of the Maritime Continent. The thermocline depth associated with the cold WPMM phase features strong positive anomalies in the western Pacific region collocated with the anomalous easterlies, but also exhibits appreciable negative and positive anomalies in the central and eastern Pacific, respectively. Together, these patterns create a flatter meridional thermocline profile, and such conditions have been shown to correlate with periods of stronger ENSO variability \citep[][]{KirtmanSchopf98,KleemanEtAl99,FedorovPhilander00,RodgersEtAl04}. In addition, the cold ENSO phase features a prominent cyclonic circulation over the Indian Ocean (centered at $ \sim 20^\circ$S and $ \sim 90^\circ$E) which strengthens the western Walker cell and advects warm, moist air from the Maritime Continent towards northern Australia. These conditions are favorable to the Australian monsoon, and thus may explain the observed TBO strengthening during cold WPMM phases.    

Besides the tropical Pacific Ocean, the WPMM has a coherent atmospheric circulation signature in the Indian and Southern Oceans. In the Indian Ocean, the cold phase of the WPMM is associated with a prominent anticyclonic gyre centered in the central-eastern Indian ocean at $ \sim 20^\circ $S latitudes. Advection from this gyre is consistent with the dipole of SST anomalies in the Indian Ocean featuring positive (negative) SST anomalies in the eastern (western) part of the basin. In the South Pacific and south Indian Ocean, the WPMM creates a coherent pattern of anomalous easterlies spanning the full meridional extent of the Indo-Pacific basin.  

\subsection{Decadal precipitation over Australia}

The precipitation composites in Fig.~\ref{figWPMM}(d) suggest that the WPMM plays an important modulating role in the multidecadal variability of Australian precipitation as simulated by CCSM4. To assess the strength of this relationship we examine the 11-year running average of spatially averaged precipitation data over Australia computed from the 1300 y CCSM4 control run.  Figure~\ref{figPrecip} shows the resulting millennial time series, $P_\text{Australia}$,  along with the temporal pattern associated with the WPMM. As with  the WPMM, pronounced fluctuations of $P_\text{Australia}$ occur on multidecadal timescales, and the correlation coefficient between the two time series is $-0.62$. In particular, negative phases of the WPMM (i.e., negative SST anomalies over the western tropical Pacific and positive SST anomalies to the south of Australia) are associated with anomalously wet periods that can last almost a century, and positive phases result in prolonged droughts. For comparison, the correlation coefficient between $P_\text{Australia}$ and the temporal pattern of the IPO (not shown here), obtained through the first  EOF of 13 y running mean of CCSM4 Indo-Pacific SST data after linear detrending to remove model drift, is 0.44. 

As shown in Fig.~\ref{figWPMM}(d), the cold WPMM phase is associated with precipitation enhancement over the whole Australian continent. In the north, this convective intensification is dynamically consistent with the surface atmospheric winds entraining anomalously warm and moist air, which is driven primarily by subsiding circulation and divergent air motion centered over the anomalously cold western Pacific SST. Moreover, additional enhancement of precipitation can be attributed to thermodynamic factors such as positive SST anomalies off the Australian coast that surround the whole continent and are particularly pronounced over its southern part. Of course, given the deficiencies of global climate models in simulating precipitation, we cannot claim that the modulating relationships between the WPMM and decadal Australian precipitation identified in CCSM4 also apply in nature. Nevertheless, these results demonstrate that at least for the climate simulated by CCSM4 the WPMM is a physically meaningful mode, and provide motivation for seeking analogous patterns and mechanisms in the real climate system.    

\section{\label{secSummary}Concluding remarks}

In this paper, we have employed a family of modes identified via NLSA in Part~I of this work \citep[][]{SlawinskaGiannakis16} to study various aspects of coupled atmosphere-ocean variability in the Indo-Pacific sector of model and observational datasets on interannual to decadal timescales. One of our main objectives has been to study the SST, surface wind, and thermocline depth patterns associated with a family of ENSO modes and combination modes between ENSO and the annual cycle (denoted ENSO-A modes), which were recovered in Part I from unprocessed Indo-Pacific SST data. We found that in both model and observational data these patterns are in good agreement with mechanisms proposed for the phase synchronization between ENSO and the seasonal cycle \citep[][]{SteinEtAl11,SteinEtAl14,StueckerEtAl13,StueckerEtAl15,McGregorEtAl12}, whereby quadratic nonlinearities in the governing equations for the coupled atmosphere-ocean system lead to a seasonally-dependent southward shift of zonal winds following the boreal winter peak of El Ni\~no (La Ni\~na) events, generating an upwelling oceanic response restoring the ENSO state to neutral or La Ni\~na (El Ni\~no) conditions. We also showed that the surface wind patterns associated with the ENSO and ENSO-A modes produce a coupling between the Indian and Pacific ocean SST with SST patterns resembling the IOD. This relationship is purely deterministic, suggesting that the ENSO and ENSO-A modes are good predictors for IOD variability.      

Besides ENSO, we studied the SST, surface wind, and precipitation patterns associated with a family of biennial modes and biennial-annual combination modes representing the TBO in CCSM4. We showed that despite the small SST anomalies involved (about an order of magnitude smaller than typical ENSO amplitudes), these patterns are consistent with coupled atmosphere-ocean mechanisms for biennial variability of the Asian-Australian monsoon \citep[][]{MeehlArblaster02}, whereby an anomalously week Australian monsoon in boreal winter tends to be followed by anomalously strong Indian and Australian monsoons in the ensuing boreal summer and winter. Given that our TBO modes were recovered without bandpass filtering the input data, these results support the existence of the TBO as an intrinsic, physically meaningful process that can be consistently inferred from SST data alone (at least in CCSM4), although our analysis does not address the causality relationships between it and other modes of climate variability such as ENSO \citep[][]{Goswami95,MeehlArblaster02,LiEtAl06,MeehlArblaster11}. In Part~I,  we were less successful in identifying coherent TBO patterns in model and observational SST datasets other than the millenial control integration of CCSM4 studied here \citep[corroborating similar difficulties faced by other studies; e.g.,][]{StueckerEtAl15b}, but the encouraging results from CCSM4 suggest expanding this study to other datasets and/or including additional physical variables that would improve the detectability of TBO patterns.      

In the decadal timescale, we studied amplitude modulation relationships between the WPMM (the dominant mode of Indo-Pacific multidecadal variability recovered by NLSA from CCSM4 data in Part~I, featuring a prominent cluster of SST anomalies in the west-central Pacific) with ENSO, the TBO, and decadal precipitation over Australia. We found that the WPMM exhibits a significant anticorrelation with the decadal envelope of ENSO and the TBO (at the $-0.63$ and $ -0.52 $ level, respectively), with cold (warm) WPMM phases corresponding to decadal periods of enhanced (suppressed) ENSO activity. In particular, we showed that cold WPMM periods are associated with anomalous westerlies in the central equatorial Pacific and anomalously deep thermocline in the eastern Pacific, leading to a background configuration which previous studies have shown to favor ENSO activity and predictability \citep[][]{KirtmanSchopf98,KleemanEtAl99,FedorovPhilander00,RodgersEtAl04}. We also demonstrated that cold WPMM phases can favor Australian monsoon activity through anomalous westerlies in the Indian Ocean strengthening the western Walker Cell, which may explain the enhanced TBO activity during those periods.  Moreover, we showed that the WPMM anticorrelates significantly (at the -0.6 level) with decadal precipitation over Australia, and justified this correlation on the basis of advection of warm moist air over the Australian continent by the atmospheric circulation patterns associated with the WPMM.  Thus, at least in the model data studied here, the WPMM is a good predictor of prolonged droughts and/or anomalously wet periods there. Overall, even though we were not able to robustly recover the WPMM from observational data (in Part~I, we only found tentative evidence for this mode in industrial-era HadISST data), our results from CCSM4 allude to it being an important pattern of Indo-Pacific variability which could potentially be useful in conjunction with the IPO to characterize the decadal behavior of the real climate.       

In summary, the family of NLSA modes identified in this work provide an attractive low-dimensional basis to characterize numerous features of the Indo-Pacific climate system. Potential future applications of these modes include objective and accurate ways of indexing oceanic oscillations across multiple timescales, studies of statistical  and causal dependencies and their underlying physical mechanisms in the coupled atmosphere-ocean system, as well predictive models based on these modes to assess future climate fluctuations from interannual to multidecadal horizons. From a data analysis standpoint this work has demonstrated that nonlinear kernel methods, appropriately designed to take dynamics into account, can potentially overcome some of the limitations of classical linear techniques in revealing intrinsic modes of multidecadal climate variability, but our study is certainly only an initial step in this direction.  Further assessment and refinement of the patterns identified here should be a fruitful avenue for future work.

\acknowledgments
J.\ Slawinska received support from the Center for Prototype Climate Modeling at NYU Abu Dhabi and NSF grant AGS-1430051.  The research of D.\ Giannakis is supported by ONR Grant N00014-14-0150, ONR MURI Grant 25-74200-F7112, and NSF Grant DMS-1521775. We thank Sulian Thual for stimulating conversations and three anonymous reviewers for comments that helped significantly improve the paper. This research was partially carried out on the high performance computing resources at New York University Abu Dhabi. 

\bibliographystyle{ametsoc2014}

\begin{figure}[t]
\noindent\includegraphics[width=\linewidth]{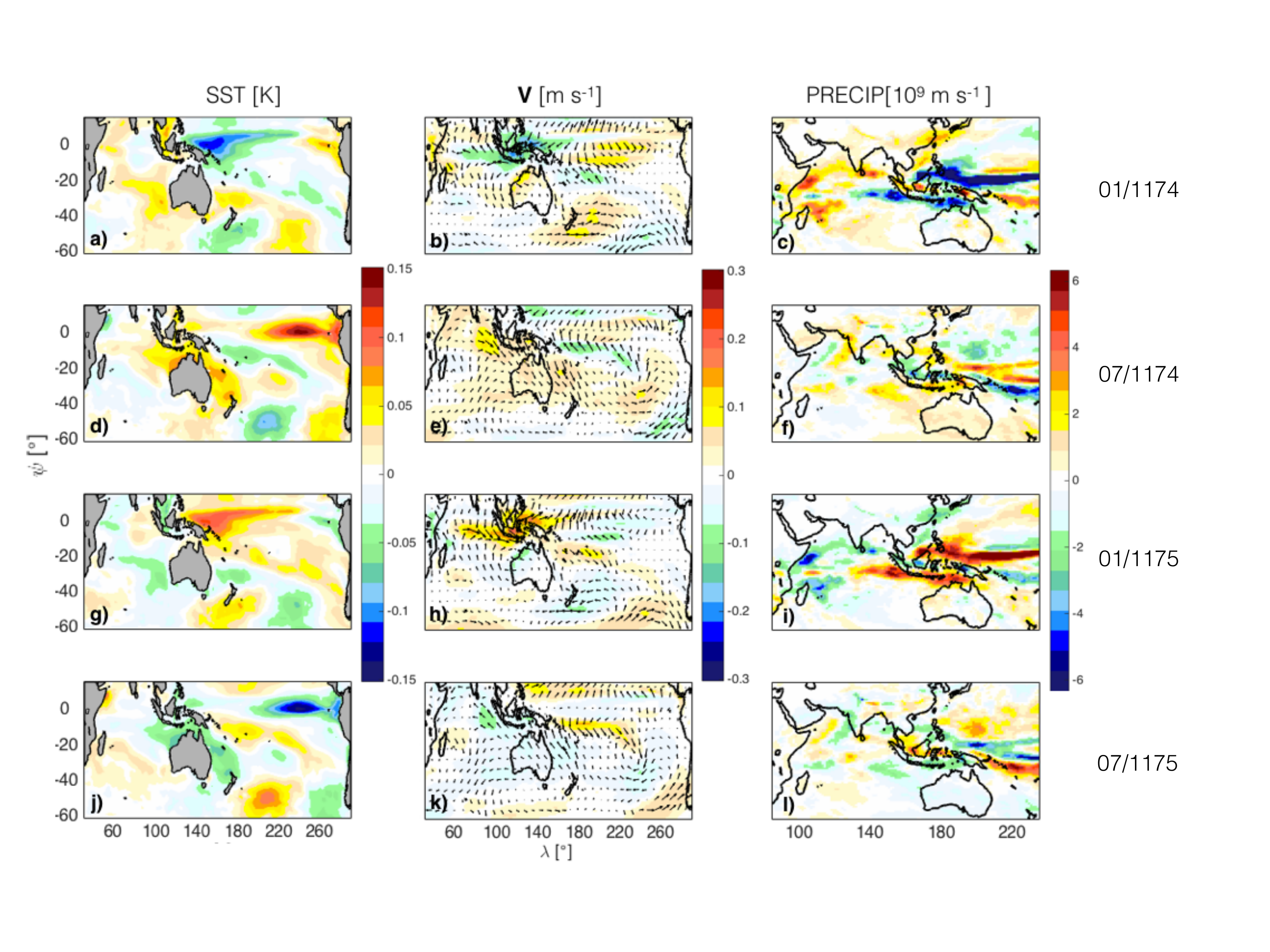}
  \caption{\label{figTBO}Snapshots of reconstructed SST (a, d, g, j), surface wind (b, e, h, k) and precipitation (c, f, i, l) anomalies based on the TBO and TBO-A mode families extracted from CCSM4 data via NLSA. The snapshots are taken every six months in the simulation period 01/1174--07/1175 and illustrate one cycle of the TBO with (a--c) an anomalously weak Australian Monsoon in January 1174, (d--f) an anomalously strong Indian Monsoon in July 1174, and (j--l) an anomalously strong Australian monsoon in July 1975.  This process is more directly visualized in movie~3 in the supplementary material.}
\end{figure}

\begin{figure}[t]
 \noindent\includegraphics[width=\linewidth]{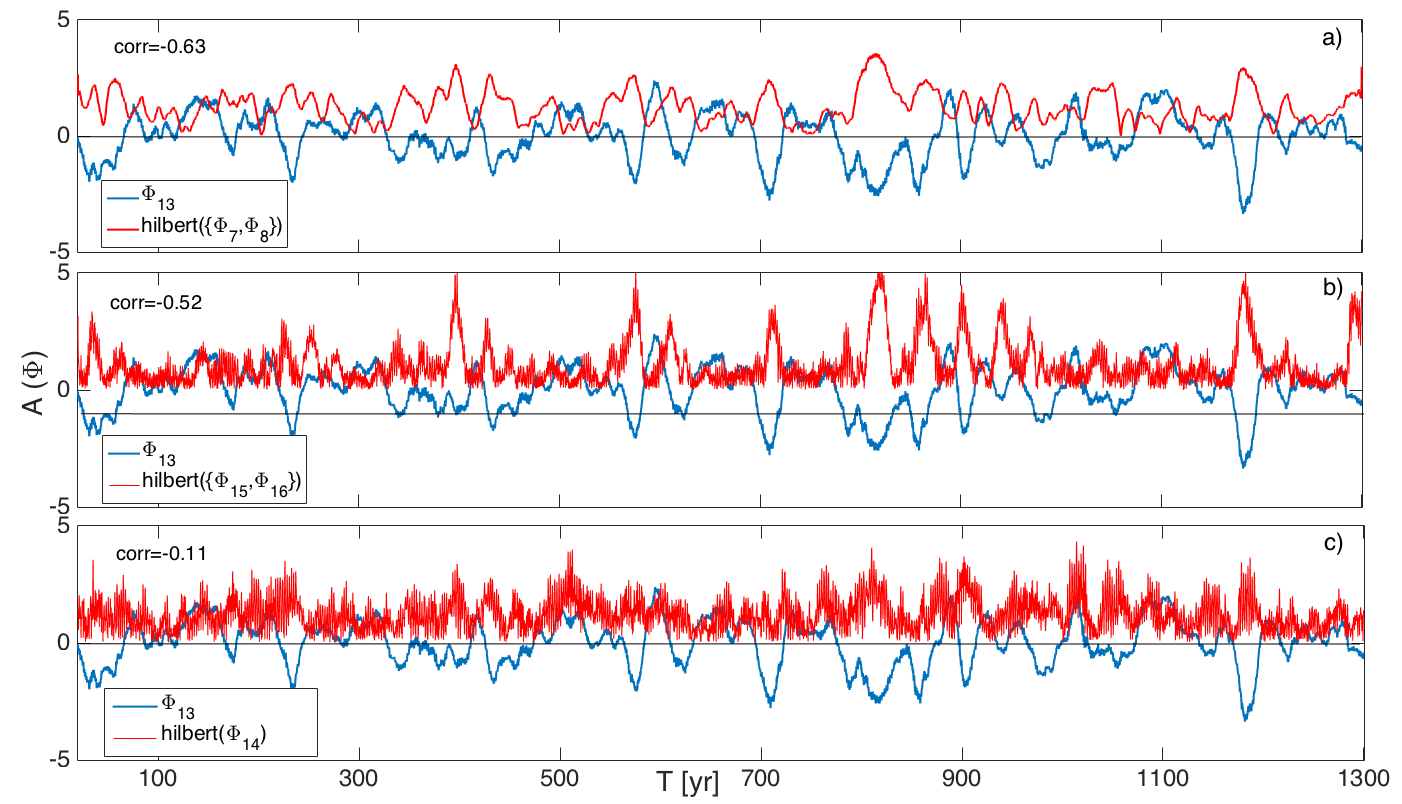}\\
 \caption{Temporal pattern of the WPMM (blue lines) and amplitude envelopes (red lines) of (a) the fundamental ENSO modes, (b) the fundamental TBO modes, and (c) the IPO mode recovered from CCSM4 data via NLSA. The amplitude envelopes were computed via the Hilbert transform of the corresponding pairs of modes. The correlation coefficients $ r $ between the plotted time series in each panel and the corresponding $ p $-values are (a) $ r = - 0.63 $, $ p \approx 0 $, (b) $ r = - 0.52 $, $ p \approx 0 $, (c) $ r = 0.11 $, $ p = 0.2 $. Here, $ p $-values were computed using a $ t $-test with  $(\text{1300 y}) \times \nu_\text{ENSO} - 2 = 323 $ degrees of freedom, representing the number of ``independent'' ENSO  events at a frequency $ \nu_\text{ENSO} = 0.25 $ y$^{-1}$ over the 1300 y CCSM4 dataset. The notation $ p \approx 0 $ means that $ p $ is numerically equal to zero for this number of degrees of freedom.}\label{figHilbert}
\end{figure}

\begin{figure}[t]
\noindent\includegraphics[width=\linewidth]{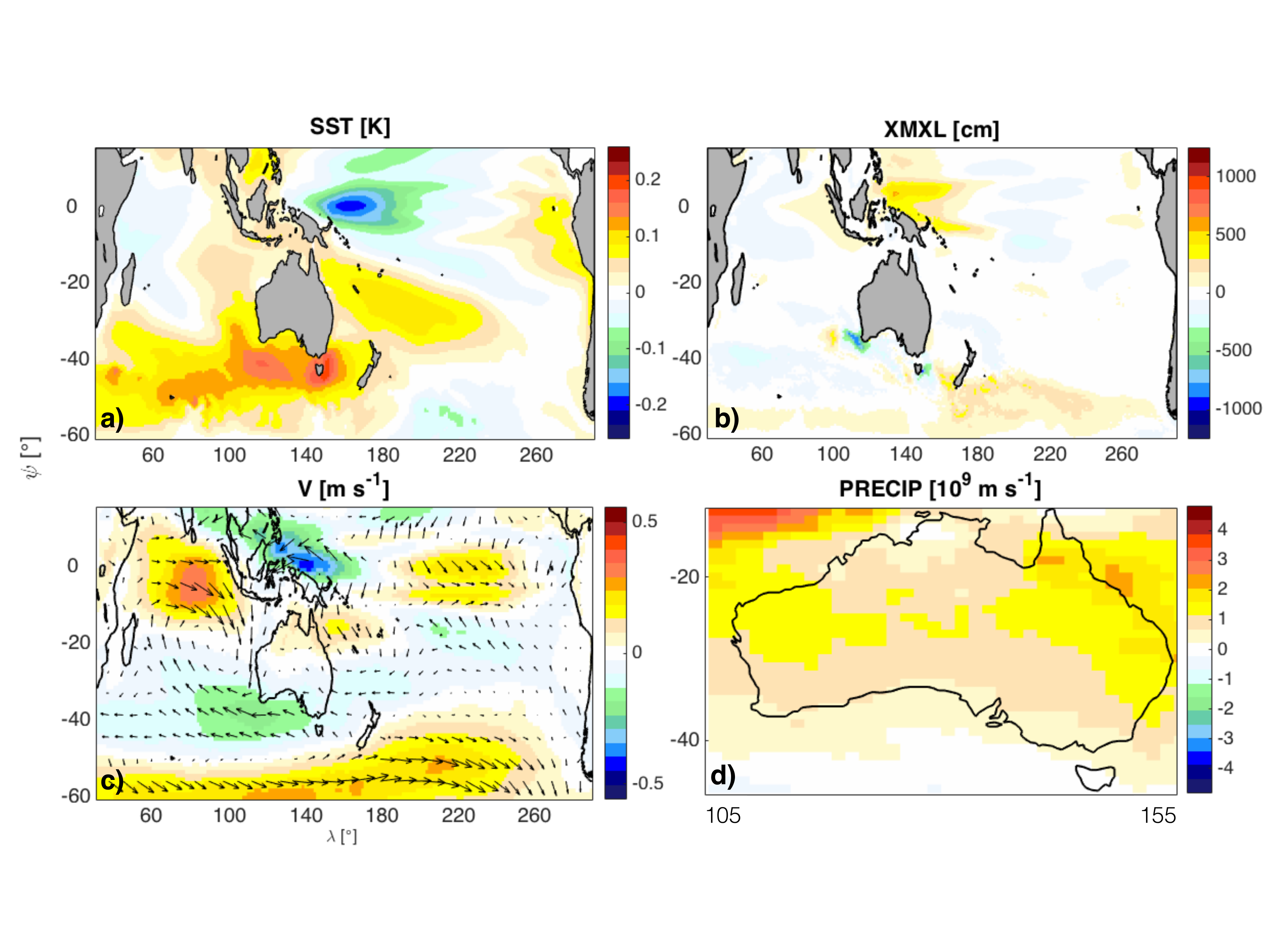}
 \caption{Composites of SST (a), maximum mixed layer depth (b), surface winds (c), and precipitation over Australia (d) based on the WPMM recovered from CCSM4 data via NLSA. The composites were constructed by averaging the corresponding reconstructed fields over the cold WPMM event (negative western Pacific SST anomalies) taking place in simulation years 1165--1185. See movie~4 in the supplementary material for a direct visualization of this period. The anomalous westerlies in the central Pacific (c) and the anomalously shallow (deep) thermocline in the central (eastern) equatorial Pacific (b) create a background known to correlate with strong ENSO activity. Similarly, the prominent cyclonic circulation in the Indian Ocean and surface wind convergence to the north of Australia (c), creates an environment favoring strong Australian monsoons, thus strengthening the TBO.}\label{figWPMM}
\end{figure}

\begin{figure}[t]
\noindent\includegraphics[width=\linewidth]{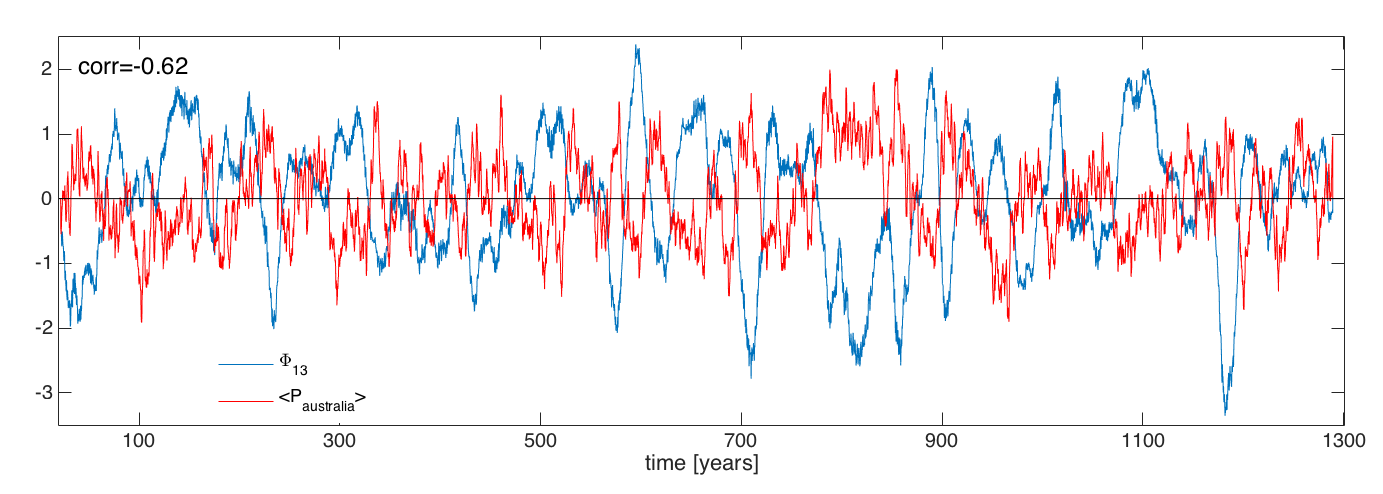}\\
 \caption{Temporal pattern of the WPMM  (blue line)  and 11-year running mean of precipitation averaged over the Australian continent (red line). The correlation coefficient of the two time series is equal to $-0.62$ at $ p $-value numerically equal to zero based on the same $ t $-test as in Fig.~\ref{figHilbert}.}\label{figPrecip}
\end{figure}

\end{document}